\documentclass[nofootinbib, floatfix, reprint,  notitlepage, longbibliography]{revtex4-1}
\usepackage{standalone}
\usepackage{amsmath}
\usepackage{amssymb}
\usepackage{graphicx}
\usepackage{verbatim}
\usepackage{xcolor}
\usepackage{import}
\usepackage{ifthen}
\AtBeginDocument{\usepackage{booktabs}}
\usepackage[pdftex, breaklinks, colorlinks, citecolor=blue, urlcolor=blue]{hyperref}

\graphicspath{{./Figs/}}

\newcommand{\ket}[1]{\left\vert{#1}\right\rangle}

\usepackage{dsfont}

\renewcommand{\vec}[1]{\ensuremath{\text{\textbf{#1}}}} 

\begin{document}

\title{Quantifying error and leakage in an encoded Si/SiGe triple-dot qubit}
\author{R. W. Andrews}
\author{C. Jones}
\author{M. D. Reed}
\author{A. M. Jones}
\author{S. D. Ha}
\author{M. P. Jura}
\author{J. Kerckhoff}
\author{M. Levendorf} 
\author{S. Meenehan}
\author{S. T.  Merkel}
\author{A. Smith} 
\author{B. Sun}
\author{A. J. Weinstein} 
\author{M. T. Rakher}
\author{T. D. Ladd}
\author{M. G. Borselli}
\affiliation{HRL Laboratories, LLC, 3011 Malibu Canyon Rd., Malibu, CA, 90265, USA}

\maketitle

\textbf{ 
Quantum computation requires qubits that satisfy often-conflicting criteria, including scalable control and long-lasting coherence~\cite{ladd2010}.
One approach to creating a suitable qubit is to operate in an encoded subspace of several physical qubits.
Though such encoded qubits may be particularly susceptible to leakage out of their computational subspace, they can be insensitive to certain noise processes~\cite{zanardi1997, lidar1998} and can also allow logical control with a single type of entangling interaction~\cite{divincenzo2000} while maintaining favorable features of the underlying physical system. 
Here we demonstrate a qubit encoded in a subsystem of three coupled electron spins confined in gated, isotopically enhanced silicon quantum dots~\cite{divincenzo2000,eng2015}.
Using a modified ``blind'' randomized benchmarking protocol that determines both computational and leakage errors~\cite{wallman2016,wood2018}, we show that unitary operations have an average total error of $\mathbf{0.35\%}$, with $\mathbf{0.17\%}$ of that coming from leakage driven by interactions with substrate nuclear spins.
This demonstration utilizes only the voltage-controlled exchange interaction for qubit manipulation and highlights the operational benefits of encoded subsystems, heralding the realization of high-quality encoded multi-qubit operations~\cite{divincenzo2000,fong2011}.
}

Electrons trapped in silicon heterostructures have many attractive features, including very long coherence times in isotopically enriched material \cite{Tyryshkin2011,yoneda2018} and compatibility with standard fabrication techniques. 
Single-spin qubits have recently demonstrated high-fidelity RF-controlled single-qubit operations~\cite{kawakami2016,yoneda2018} and two-qubit gates using the exchange interaction~\cite{Zajac2017,Watson2018,Huang2018}.  
However, using RF signals for single-qubit control requires a large, stable magnetic field and introduces challenges with crosstalk.   
Fortunately, electron spins are particularly well-suited to forming encoded qubits. 
Two coupled electron spins can be operated at near-zero magnetic field as a ``singlet-triplet'' qubit~\cite{levy2002, petta2005}.  
That qubit is insensitive to uniform magnetic field fluctuations but still requires a magnetic field gradient for universal control.  
Three coupled electrons~\cite{russ2016} can form a qubit with a tunable electric dipole moment, which could enhance RF selectivity, 
or the exchange-only qubit, which can be universally controlled using only the exchange interaction and does not require synchronization of gate operations with a local oscillator. 
Exchange is highly local and can be accurately controlled with a large on-off ratio using only fast voltage pulses.
The combination of these features makes the exchange-only qubit especially attractive for use in many-qubit systems.

\begin{figure}[t]
\includegraphics[width=1\linewidth]{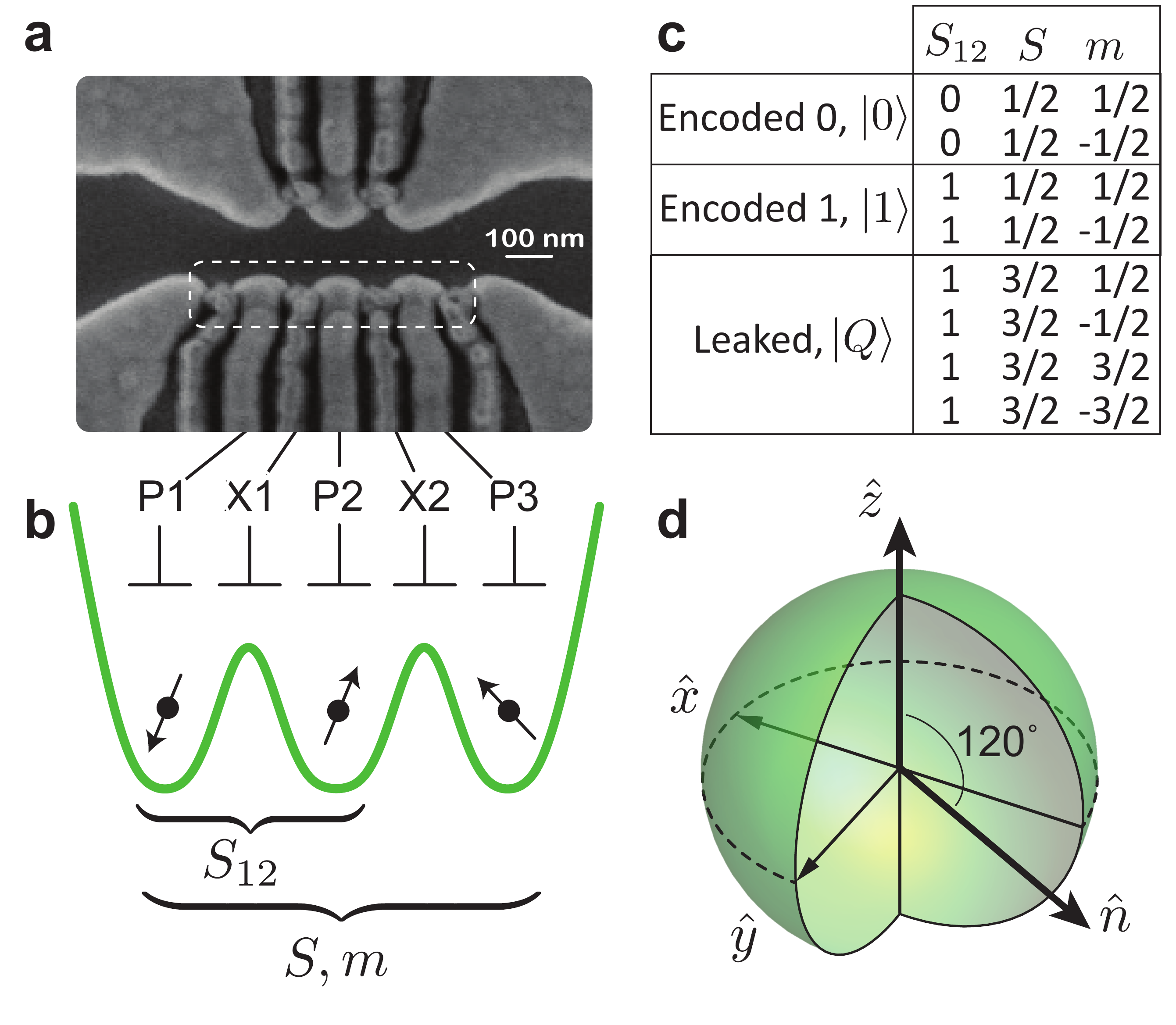}
\caption{
An encoded qubit with three electron spins.  
{\bf{a}}, Scanning electron micrograph of a device with three coupled quantum dots and a nearby dot charge sensor.  
Aluminum gates (bright in image) confine electrons in two dimensions while a Si/SiGe heterostructure provides out-of-plane confinement.  
Quantum dot array is formed beneath the region indicated by the dashed box.  
{\bf{b}}, Cartoon of potential energy landscape along the quantum dot array.  
The gates labelled `X1' and `X2' predominantly control the potential barrier between neighboring electrons and `P1' through `P3' the associated dot chemical potential.
We manipulate the encoded qubit state by modulating the voltage applied to these gates.  
{\bf{c}}, Three-electron states in the coupled-spin basis.  
Initializing the qubit state in $\ket{0}$ involves preparing the two electrons confined by P1 and P2 in a spin singlet while leaving the third electron uninitialized.
{\bf{d}}, In the Bloch sphere representation, the electron-electron interaction modulated by the potential barrier under the `X1' gate induces a qubit rotation about the $\hat{z}$ axis; `X2' similary induces a rotation about the  $\hat{n}$ axis, which is separated from the $\hat{z}$ axis by 120 degrees. 
}
\label{fig1}
\end{figure}

Here we demonstrate high-fidelity operation of a triple-dot, exchange-only qubit.
Electrons are trapped in quantum dots formed by the electrostatic potentials of a Si/SiGe heterostructure and the patterned metallic gates shown in Fig.~\ref{fig1}a \cite{eng2015}.  
As depicted in Fig.~\ref{fig1}b, we manipulate the chemical potential of the dots by primarily adjusting voltages applied to gates P1, P2, and P3 and the exchange coupling between neighboring electrons by adjusting voltages applied to gates X1 and X2.  
The three coupled electrons constitute an 8-dimensional spin system that can be used to form an exchange-only qubit \cite{divincenzo2000, russ2016}.   

Operation of this encoded qubit is conveniently described in terms of the coupled spin states $\ket{S_{12}, S, m}$, where $S$ is total spin quantum number, $S_{12}$ is the total spin of the electrons confined beneath gates P1 and P2, and $m$ is the spin projection in an arbitrary direction, such as that of an applied magnetic field.
The 8 possible states are enumerated in Fig.~\ref{fig1}c. 
We use energy-selective initialization~\cite{petta2005, eng2015} to prepare the first two electrons in a spin singlet state, giving $S_{12}$=0.  
Since the spin state of the third electron remains random, the qubit is initialized in an incoherent mixture of $\ket{0, \frac{1}{2}, \frac{1}{2}}$ and $\ket{0, \frac{1}{2}, -\frac{1}{2}}$.  
These states collectively comprise encoded 0 (denoted hereafter as $\ket{0}$). 
In this encoding, the quantum number $m$ does not affect logical operations and the qubit is insensitive to global magnetic fields.  
We measure the $S_{12}$ quantum number (which distinguishes between $\ket{0}$ and the six other spin states) via spin-to-charge conversion~\cite{petta2005,eng2015} and charge-state detection using a dot charge sensor (formed using gates whose leads are visible in the top of Fig.~\ref{fig1}a) and associated amplification circuitry \cite{eng2015,jones2018}.  
We denote $P_0$ as the projector onto $S_{12}=0$ and the probability of being found in $\ket{0}$ for a density matrix $\rho$ as $\mathrm{Tr}(P_0 \rho)$.  
More detail on this subsystem algebra appear in the Supplemental Information. 

To manipulate the encoded qubit state, we induce interaction between neighboring pairs of electrons by applying voltage pulses to device gates in order to lower the inter-dot potential barrier (see Fig.~\ref{fig1}b).  
The resulting electron-electron exchange interaction only affects the $S_{12}$ quantum number \cite{loss1998, divincenzo2000, eng2015}, so only states within the $S=1/2$ subspace are accessible through exchange. 
However, other physical mechanisms may alter the total spin and drive the qubit into leaked states with $S=3/2$; in our case, leakage is generally caused by magnetic gradients such as those due to hyperfine interactions with nuclear spins.    
In the Bloch sphere representation, exchange between the two pairs of electrons induces rotations about non-orthogonal axes separated by 120 degrees \cite{divincenzo2000, ladd2012, russ2016}.  
As diagrammed in Fig.~\ref{fig1}d, we label them as $\hat{z}$ and $\hat{n}$.

\begin{figure}[h]
\includegraphics[width=1\linewidth]{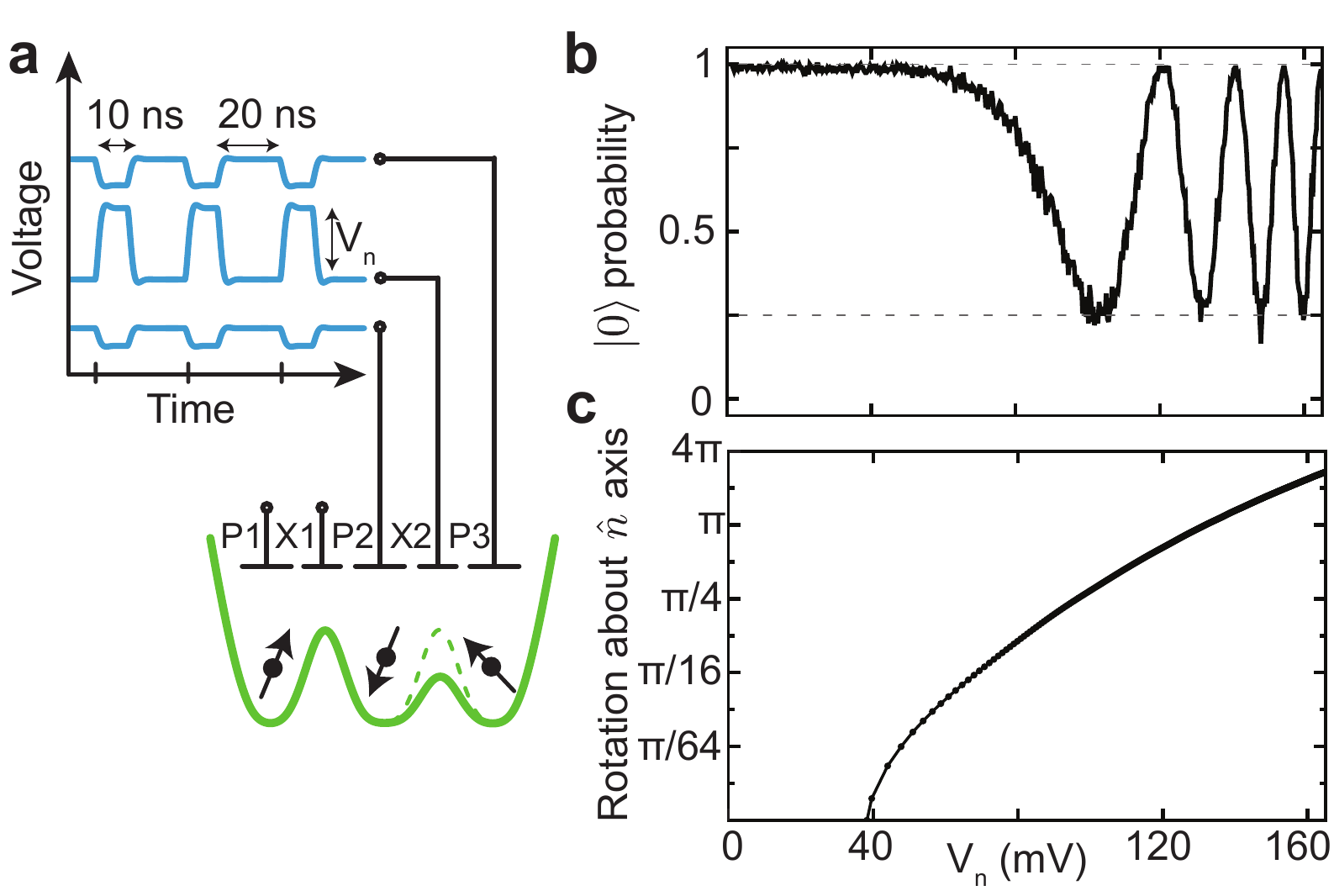}
\caption{
Voltage-to-angle calibration of the $\hat{n}$ axis.  
{\bf{a}}, To perform our calibration sequence, the qubit is initialized in $\ket{0}$ and then three voltage pulses are applied to the P2, X2, and P3 gates with the relative pulse amplitudes and durations specified in the main text. 
Voltage pulses are drawn to highlight non-idealities arising from control bandwidth limitations.  
After rotation, the qubit is measured.  
{\bf{b}}, Measured probability of finding the qubit in $\left|0\right\rangle$ ($S_{12}$=0) after applied voltage pulses.  
Voltage pulses cause the qubit to rotate about the $\hat{n}$ axis, leading to a maximum expected contrast of 0.75 (indicated by dashed lines).
{\bf{c}}, By unwrapping the accrued phase in sinusoidal waveforms like that in {\bf{b}}, we create a map between applied voltage and rotation from a single pulse (see Supplement for details).
}
\label{fig2}
\end{figure}

To demonstrate high-fidelity encoded operations, we first create a mapping between the voltage of the pulses we apply and the resulting exchange rotation.  
A rotation about the $\hat{n}$ axis involves voltage pulses on the P2, X2, and P3 gates, labelled in Fig.~\ref{fig1}b.  
We choose the ratios of the voltages applied to each of these gates so that we primarily alter the potential barrier between the electrons while minimizing differences in the chemical potentials of the dots (see Fig.~\ref{fig2}a).  
This ``symmetric operation'' minimizes sensitivity to charge noise \cite{reed2016, martins2016}.  
The relative gate voltages required for symmetric operation can be found using a procedure outlined in \cite{reed2016}. 
For $\hat{n}$ rotations in this device, we pulse the voltage along the vector $\vec{V}(V_{n}) =\vec{V}_0+ V_{n}\left\{ -0.20, 1, -0.25  \right\}$ in the $\left\{ \mathrm{P2}, \mathrm{X2}, \mathrm{P3}\right\}$ voltage coordinates.   
Between pulses, gates are biased at an ``idle'' configuration $\vec{V}_0$ in which all electrons are isolated and exchange rates are negligible.  
We map out qubit rotation by intializing the qubit, applying a sequence of three consecutive voltage pulses that are 10~ns wide and separated by an idle time of 20~ns (diagrammed in Fig.~\ref{fig2}a), and measuring the resulting probability of finding the qubit in $\ket{0}$ (Fig.~\ref{fig2}b).  
For this device, a single-pulse $2\pi$ rotation about the $\hat{n}$ axis occurs at $V_{n}\approx150$ mV.  
By appropriately processing the data in Fig.~\ref{fig2}b (see Supplemental Information), we create a voltage-to-rotation mapping for a single pulse, as shown in Fig.~\ref{fig2}c.  
We see from this calibration that an increase in $V_{n}$ by 120 mV changes the resulting rotation angle by more than 3 orders of magnitude, exemplifying the large dynamic range available with this control technique.  
A similar procedure is used for mapping rotations about the $\hat{z}$ axis.   

\begin{figure}[h]
\includegraphics[width=1\linewidth]{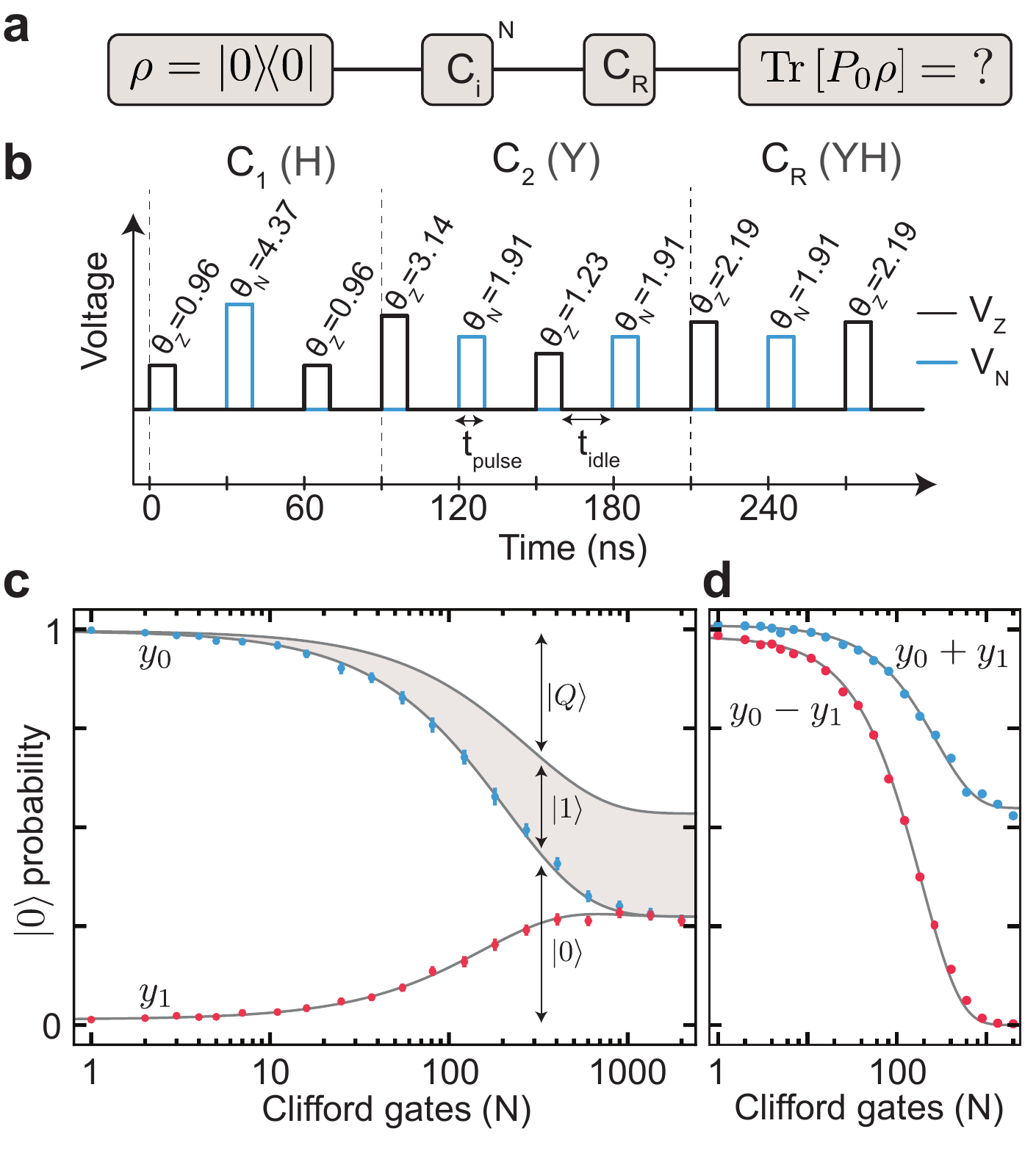}
\caption{
Randomized benchmarking of an encoded qubit.   
{\bf{a}}, Benchmarking involves initializing the qubit ($\rho=|0\rangle\!\langle0|$), applying a sequence of $N$ gates drawn randomly from the Clifford group ($C_{i}^{N}$), applying a recovery operation ($C_{R}$), and measuring the result.  
{\bf{b}}, Example pulse sequence for $N=2$.  
Qubit control is performed with interleaved voltage pulses that induce rotations about the $\hat{n}$ and $\hat{z}$ axes.
{\bf{c}}, $|0\rangle$ probability versus sequence length.  
For each sequence, the recovery Clifford is randomly selected to ideally return the state to either $\ket{0}$ ($y_0$ data shown in blue) or $\ket{1}$ ($y_1$ data in red). 
Error bars indicate the standard error of the mean probability from 60 sequence repetitions, and when not visible are smaller than the size of data points (see Methods).
Vertical arrows label curves denoting the estimated population in states $\ket{0}$ (coinciding with $y_0$ fit), $\ket{1}$ (the shaded area between $y_0$ and the curve above), and $\ket{Q}$ (the remaining population) as a function of $N$.
See Methods for additional data acquisition details.
{\bf{d}}, Single-exponential fits to sum and difference data traces. 
}
\label{fig3}
\end{figure}

Calibrated rotations about the two axes can be interleaved to generate arbitrary single-qubit operations \cite{lowenthal1971}.  
For example, an encoded Hadamard gate is applied with a sequence of a ~0.96 radian rotation about the $\hat{z}$ axis, a ~4.37 radian rotation about the $\hat{n}$ axis, and then a second ~0.96 rotation about the $\hat{z}$ axis; this type of sequence is diagrammed in Fig~\ref{fig3}b. 
For the construction used here, the 24 quantum gates in the Clifford group require an average of 2.7 individual rotations about some combination of the $\hat{n}$ and $\hat{z}$ axes, with the shortest Clifford requiring one rotation and the longest four.  
A complete table appears in the Supplementary Information.
This construction is less efficient than possible with systems that have orthogonal rotation axes, where Clifford operations require an average of 1.9 rotations~\cite{kawakami2016}. 
The advantage of the triple-quantum-dot, however, lies in its simplicity: all operations only require nearest-neighbor voltage-controlled exchange interactions.  
Two-qubit gates can also be implemented using only exchange between two neighboring triple-dot encoded qubits~\cite{divincenzo2000,fong2011}.

We demonstrate high-fidelity operation of this encoded qubit using a modified form of randomized benchmarking (RB)~\cite{Emerson2007,knill2008}. 
A conventional RB experiment (diagrammed in Fig.~\ref{fig3}a and b) consists of initializing the qubit to $|0\rangle$,  applying a series of $N$ gates randomly chosen from the Clifford group, applying a final ``recovery'' operation $C_R$ that ideally inverts the preceding sequence, and measuring the resulting $y_0(N) = \mathrm{Tr}[P_0 \rho(N)]$, where $\rho(N)$ is the qubit density matrix after $N$ gates and the recovery gate.  
The average gate error rate $\varepsilon$ is determined by sweeping $N$ and fitting $A + B(1-2\varepsilon)^N$ to $y_0$.
Data resulting from such an experiment are shown as blue data points in Fig.~\ref{fig3}c.  
These data reveal that leakage plays a significant role in qubit operation: The $|0\rangle$ probability approaches 0.25 after many Clifford operations, indicating that the final state is a uniform mixture of the 8 three-electron spin states.  

We extend conventional RB to measure both qubit errors and leakage errors by exploiting two features of our qubit.  First, all leaked states have $S_{12} = 1$ and so yield the same measurement outcome as $\ket{1}$.  
Second, exchange operations preserve $S$ and do not cause leakage.
For each RB sequence, we randomly select the recovery operation to either return the qubit to $\ket{0}$ (as in conventional RB) or $\ket{1}$, keeping track of which recovery was used while still making the experiment ``blind'' to the correct final state.  
Similar modifications to the recovery operation have been studied in other leakage-detecting variants of RB~\cite{wallman2016,helsen2017,wood2018, xue2018}.
The measurement probabilities as a function of $N$ are binned in two groups $y_0$ and $y_1$, where the subscript denotes the selection of recovery operation, as shown in Fig.~\ref{fig3}c.  
We extract error and leakage using the ansatz (see Supplemental Information for details)
\begin{align}
y_0 = A + B(1-p)^N + C(1-q)^N, \label{eqn::y0} \\
y_1 = A - B(1-p)^N + C(1-q)^N. \label{eqn::y1}
\end{align}  
The problem of fitting a coupled system of equations is made simpler by separately fitting single exponentials to $y_0 + y_1$ and $y_0 - y_1$ to estimate parameters $A, B, C, p$ and $q$ (see Fig.~\ref{fig3}d).

To define error and leakage per gate, we must re-examine the effect of state preparation and measurement (SPAM) errors.  
In conventional RB, the dimensionality of the system is an assumption (e.g. $d=2$ for one qubit) that becomes a normalization factor in determining gate error.  
With leakage, the situation is more complicated because the number of accessible leaked states might not be known.  
Moreover, if there are multiple leaked states, then population transfer to these states may occur at different rates since only the qubit subsystem undergoes Clifford twirling to depolarize errors.
To address these complications, we define the leakage error $\Gamma$ and total error $\varepsilon$ using the decay rates $p$ and $q$, as well as parameters $B$ and $C$ that are typically associated with SPAM (see Supplemental Information for complete derivation).  
We assign total error to all of the population leaving subspace $P_0$ in the first gate, normalized by SPAM fidelity $2B$:
\begin{equation}
\varepsilon = \mathrm{Tr}\left\{P_0[\rho(0) - \rho(1)]\right\}/(2B) = p/2 +Cq/(2B).
\label{eqn::total_error}
\end{equation}
If there is no leakage, $C = 0$ and $\varepsilon$ coincides with the gate error from conventional RB.  
The leakage probability is likewise defined as the total population leaking from the encoded qubit subspace after the first gate, again normalized by $2B$:
\begin{equation}
\Gamma = \mathrm{Tr}\left\{[P_L[\rho(0) - \rho(1)]\right\}/(2B) = Cq/B,
\label{eqn::leak_error}
\end{equation}
where $P_L$ is the projector onto leaked states.  
Applying this analysis to the data depicted in Fig.~\ref{fig3}c, we obtain a total error rate $\varepsilon = 0.35\%$, where leakage $\Gamma = 0.17\%$ contributes approximately half of the error.
We identify the non-leak error $(\varepsilon - \Gamma)$ as the ``qubit error.'' 
We also extract a SPAM fidelity $F_{\mathrm{SPAM}} = 0.5+B = 99.2\%$.

\begin{figure}[h]
\includegraphics[width=1\linewidth]{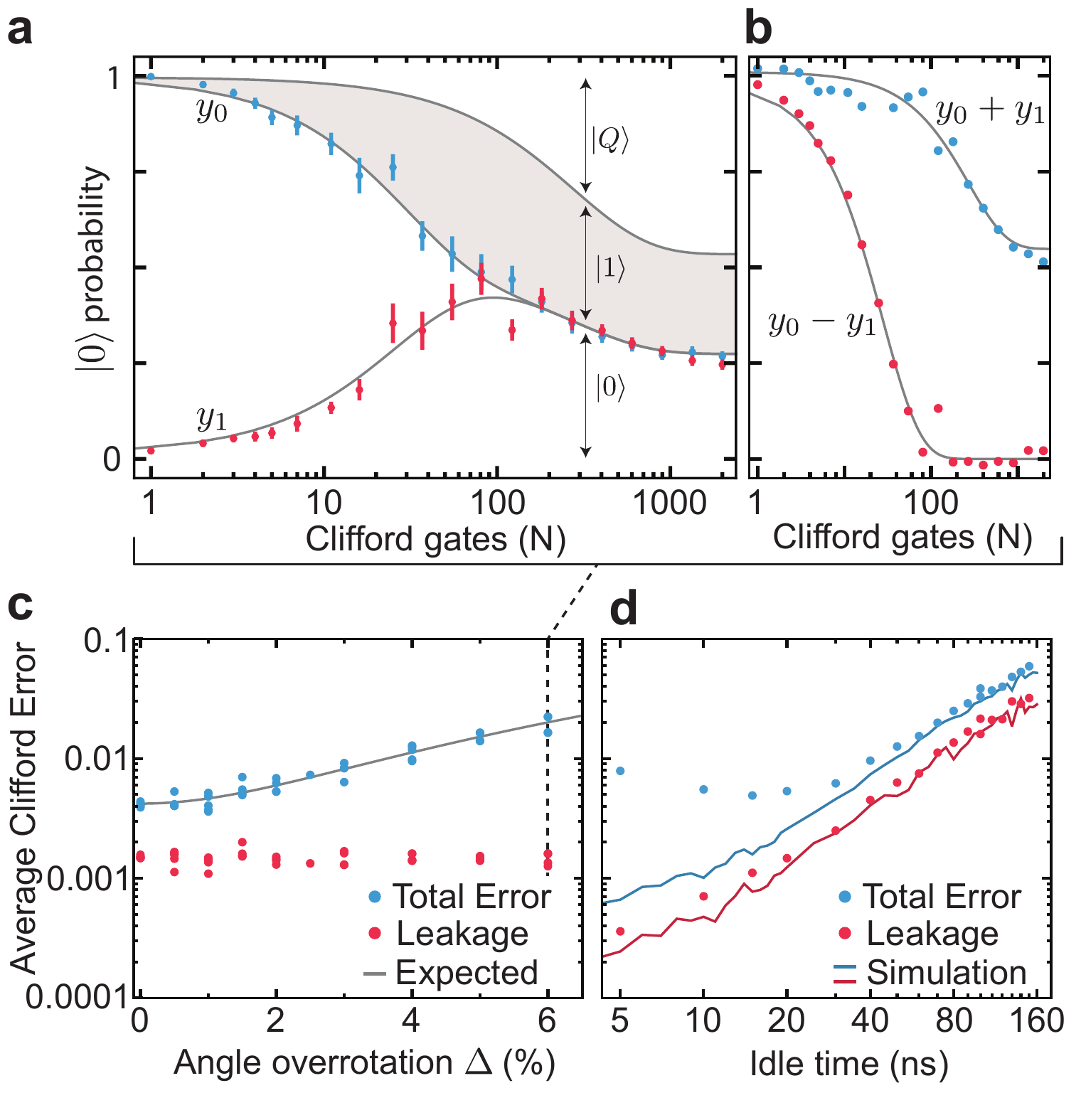}
\caption{
Blind RB versus overrotation and idle time.  
{\bf{a}}, Blind RB dataset for $\Delta = 6\%$ overrotation, showing non-exponential decay due to substantially different rates of  qubit and leakage error.   
Error bars indicate the standard error of the mean probability from 60 sequence repetitions, and when not visible are smaller than the size of data points (see Methods). 
{\bf{b}}, Sum and difference combinations of $\Delta = 6\%$ data each show single-exponential decay.  
{\bf{c}}, Total error per Clifford (blue) and leakage (red) for increasing angle overrotation. 
The angles used to construct the Clifford group were altered by a fixed ratio $\Delta$, with the sign randomly chosen to provide an over- or under-rotation (see main text). 
Multiple points at each value of $\Delta$ denote distinct sets of randomly generated signs of overrotation. 
All data were taken with 10 ns pulse time and 20 ns idle time. 
Line shows expected scaling of $2.7(\pi^2/6)\Delta^2 \approx 4.4\Delta^2$ with an offset of 0.0042 (see Supplemental Information) .  
{\bf{d}}, Total error and leakage for varying idle times (shown as $t_{\mathrm{idle}}$ in Fig.~\ref{fig3}b). 
Overlaid lines show total error and leakage produced by simulations that include the effects of hyperfine magnetic fluctuations and charge noise on pulses (see Supplemental Information). 
}
\label{fig4}
\end{figure}

To confirm that blind RB correctly extracts leakage, we introduce intentional overrotation errors on the control pulses.  
Exchange preserves total spin $S=1/2$, and so control imprecision only causes qubit errors and not leakage.    
We modify each value $\theta_i$ in the set of angles used to construct the Clifford operations as $\theta_i \rightarrow (1 \pm \Delta) \theta_i$, where the sign of overrotation is randomized for each angle and $\Delta$ is fixed.  
The gate voltages necessary to produce these rotations are found using calibrations like the one shown in Fig.~\ref{fig2}c.  
We perform blind RB for values of $\Delta$ from 0\%-6\%, with each value repeated several times with different random overrotation signs.
The resulting leakage and total error rates are shown in Fig.~\ref{fig4}c.  
As expected, the total error increases with $\Delta$ but leakage remains constant at $\Gamma \approx 0.15\%$.  
We also see that blind RB reliably differentiates between qubit error and leakage error: for $\Delta = 6\%$, shown in Fig.~\ref{fig4}a and \ref{fig4}b, we resolve a leakage error per Clifford of only 0.15\% from a total error of 2.0\%, a difference which leads to a distinctly non-exponential decay in $y_0$.  

We use a combination of blind RB and spin simulations to test our understanding of the error per Clifford gate.
We perform blind RB as a function of the idle time ($t_{\mathrm{idle}}$) since hyperfine interactions are anticipated to increase both total error and leakage quadratically with $t_{\mathrm{pulse}}+t_{\mathrm{idle}}$~\cite{ladd2012}.
Fig.~\ref{fig4} shows that for $t_{\mathrm{idle}}>30$~ns, error increases as expected and the ratio of leakage to total error approaches 0.5.
However, for shorter idle times, qubit error increases relative to leakage.  
The high-$t_{\mathrm{idle}}$ behavior is well-replicated by detailed simulations of three spins subject to both charge and magnetic noise.
Charge noise is modeled as fluctuating exchange with noise power similar to that reported in Ref.~\onlinecite{reed2016} and magnetic noise as independently fluctuating vector magnetic fields at each dot consistent with contact hyperfine coupling to the 800~ppm $^{29}$Si and natural $^{73}$Ge nuclear environment.  
Both noise sources are modeled with a $1/f$ noise spectral density~\cite{eng2015} (see the Supplementary Information).
We suspect the divergence between the simulations and the data at short $t_{\mathrm{idle}}$ is due to the finite bandwidth of our voltage pulses preventing accurate rotations.
Such control errors are similar in nature to overrotation error, do not cause leakage, and become worse with shorter $t_{\mathrm{idle}}$ due to unintentional pulse overlap.
Using the same expression relating overrotation to qubit error (Fig.~\ref{fig4}c caption), the combined effects of imperfect control and charge noise can be modeled as stochastic overrotation.  For total error $\varepsilon = 0.35\%$ and leakage $\Gamma = 0.17\%$ at the minimum total error point of $t_{\mathrm{idle}}=20$~ns, the root-mean-square overrotation amount is about 1.3\%.

Despite the necessity of multi-pulse composite rotations and its susceptibility to leakage out of the encoded subspace, we have shown that our implementation of an exchange-only qubit using isotopically-enhanced silicon is capable of thousands of coherent qubit manipulations.
This demonstration, performed at a low magnetic field and requiring only baseband voltage control, bodes well for future demonstrations with larger quantum dot arrays.  
Improvements in qubit error are still required, however.  
Through its isolation of leakage error, the blind randomized benchmarking protocol shows that total error is primarily limited by a combination of dephasing from nuclear spins and control error.  
While various strategies for dynamically correcting for magnetic gradients are known~\cite{hickman2013}, the most straightforward path for reducing nuclear dephasing is faster pulsing.
Reducing the effects of control errors, particularly when operating faster, will require improvements of control bandwidth, software-based precompilation of pulses to accelerate pulse settling \cite{Johnson2010}, or a combination thereof.  
These engineering concerns will be of particular importance when tackling the longer pulse sequences required for two-qubit exchange-only gates~\cite{divincenzo2000,fong2011} and beyond.

\section*{Methods}
The device utilized here was fabricated with an overlapping aluminum gate architecture similar to previously published devices \cite{Borselli2015,Zajac2015,reed2016}. 
The device was measured in a dilution refrigerator with a base temperature of 20 mK. 
High bandwidth coaxial cables allowed for rapid pulsing of the voltages applied to gates P1, X1, P2, X2, P3, necessary for performing randomized benchmarking at the point of symmetric operation. 
Voltage pulses were sourced by arbitrary waveform generators with a 400 MHz sample rate, the outputs of which were low-pass filtered with a cutoff of 120 MHz and a flat time delay.
All experiments were performed with 10 ns pulse time and 20 ns idle time (unless indicated otherwise) and with zero applied magnetic field.

Blind randomized benchmarking was performed by generating a random sequence of Clifford gates of a given length, $N$, with the final Clifford randomly chosen to invert to $\ket{0}$ or $\ket{1}$. 
This sequence is combined with state preparation and measurement and then averaged over 100 runs before moving to the next sequence length. 
The length of the Clifford sequence is then swept sixty times, from $N=1$ to $N=2000$.  The sixty measurements for each sequence length are combined, and their average and standard deviation are the data points and error bars shown in Fig.~\ref{fig3}c. 
The same averaging conditions were used for the data of Fig.~\ref{fig4}.

Measurement of $S_{12}$ was accomplished through spin-to-charge conversion and charge sensing, with high SNR ($\sim$9) single-shot discrimination enabled by cryogenic HEMT amplification, as in Ref.~\onlinecite{jones2018}. 
Each measurement was individually thresholded to map each result to 0 or 1.
To avoid populating excited states when transitioning between gate voltages used for measurement and idle configurations, voltages on gates P1 and P2 were slewed adiabatically when near the region of spin-to-charge conversion. 
The total sequence length of state preparation, adiabatic ramping, and measurement with a single exchange pulse was 137.1~$\mu s$.

\bibliography{BlindRB_main.bbl}

\section*{Author Contributions}
The device was designed, characterized, and fabricated by S.D.H., M.P.J., M.L., M.T.R., and M.G.B.
Control theory, coding, and analysis were provided by C.J., A.M.J., S.M., S.T.M., A.S., A.J.W., and T.D.L.  
Measurements were made by R.W.A., M.D.R., A.M.J., J.K., S.M., B.S., and A.J.W.  
The manuscript was written by R.W.A., C.J., M.D.R., A.M.J, and T.D.L. with input from all authors.
The effort was supervised by M.T.R., T.D.L., and  M.G.B.

\section*{Ackowledgements}
We thank Andy Hunter, Kevin Eng, Mark Gyure, and Bryan Fong for valuable contributions leading to this work.  

\section*{Data Availability}
The data that support the findings of this study are available from the corresponding author upon reasonable request.

\section*{Competing interests}
The authors declare no competing interests.  

\end{document}


\title{Supplemental Material: Quantifying error and leakage in an encoded Si/SiGe triple-dot qubit}
\author{R. W. Andrews}
\author{C. Jones}
\author{M. D. Reed}
\author{A. M. Jones}
\author{S. D. Ha}
\author{M. P. Jura}
\author{J. Kerckhoff}
\author{M. Levendorf} 
\author{S. Meenehan}
\author{S. T.  Merkel}
\author{A. Smith} 
\author{B. Sun}
\author{A. J. Weinstein} 
\author{M. T. Rakher}
\author{T. D. Ladd}
\author{M. G. Borselli} 
\affiliation{HRL Laboratories, LLC, 3011 Malibu Canyon Rd., Malibu, CA, 90265, USA}

\maketitle

\section{Encoding Details for the Exchange-Only Qubit}
\label{sec:DFS}

\begin{wraptable}[35]{r}{0.45\textwidth} 
\vspace{-0.3in}  
\newcommand{\ta}{\theta_1}
\newcommand{\tb}{\theta_2}
\newcommand{\tc}{\theta_3}
\newcommand{\td}{\theta_4}
\caption{Composite rotation recipes employed for the 24 single-qubit Clifford gates.  $I$ is the identity; $X,Y,Z$ are Pauli operators, $S=\sqrt{Z}$, and $H$ is the Hadamard.   The table indicates the order of rotations $R_z(\theta)$ and $R_n(\theta)$ about respective Bloch-sphere axes $\hat{z}$ and \mbox{$\hat{n}=-\hat{z}/2-\sqrt{3}\hat{x}/2$} chosen to achieve each gate up to an overall phase, and the angles required.  The angles employed are \mbox{$\ta=\tan^{-1}\sqrt{8}\approx 70.529^\circ$}, 
\mbox{$\tb = \pi-\tan^{-1}(\sqrt{5}/2)\approx 131.81^\circ$}, 
\mbox{$\tc \approx 74.755^\circ$}, 
and \mbox{$\td \approx 201.625^\circ$}, 
e.g. \mbox{$SX=e^{i\delta}R_n(\td)R_z(\tb)R_n(\tc)$}.}
\label{table:Clifford_angles}
$$
\begin{array}{|c|c|c|c|c|}
\hline
\textrm{%
    Gate} & R_z         & R_n     & R_z          & R_n    \\\hline
\hline 
I          & -            & -       & -            & -      \\   
\hline 
X          & -            & \pi-\ta & \ta          & \pi-\ta \\  
Y          & \pi          & \pi-\ta & \ta          & \pi-\ta \\  
Z          & \pi          & -       & -            & -       \\  
\hline 
S          & 3\pi/2       & -       & -            & -       \\  
S^\dag     & \pi/2        & -       & -            & -       \\  
\hline 
SX         & -            & \tc     & \tb          & \td     \\  
S^\dag X   & -            & \td     & \tb          & \tc     \\  
\hline 
H          & (\pi-\ta)/2  & \pi+\ta & (\pi-\ta)/2  & -       \\  
\hline 
XH         & (\pi+\ta)/2  & \pi-\ta & (3\pi+\ta)/2 & -       \\  
YH         & (\pi+\ta)/2  & \pi-\ta & (\pi+\ta)/2  & -       \\  
ZH         & (3\pi+\ta)/2 & \pi-\ta & (\pi+\ta)/2  & -       \\  
\hline 
SH         & (\pi-\ta)/2  & \pi+\ta & 2\pi-\ta/2   & -       \\  
HS         & 2\pi-\ta/2   & \pi+\ta & (\pi-\ta)/2  & -       \\  
S^\dag H   & (3\pi+\ta)/2 & \pi-\ta & \ta/2        & -       \\  
HS^\dag    & \ta/2        & \pi-\ta & (3\pi+\ta)/2 & -       \\  
\hline 
HSH        & \ta/2        & \pi-\ta & \ta/2        & -       \\  
HS^\dag H  & \pi+\ta/2    & \pi-\ta & \pi+\ta/2    & -       \\  
S^\dag HS  & \pi+\ta/2    & \pi-\ta & \ta/2        & -       \\  
SHS^\dag   & \ta/2        & \pi-\ta & \pi+\ta/2    & -       \\  
\hline 
HSX        & \ta/2        & \pi-\ta & (\pi+\ta)/2  & -       \\  
S^\dag XH  & (\pi+\ta)/2  & \pi-\ta & \ta/2        & -       \\  
H S^\dag X & 2\pi-\ta/2   & \pi+\ta & (3\pi-\ta)/2 & -       \\  
SXH        & (3\pi-\ta)/2 & \pi+\ta & 2\pi-\ta/2   & -       \\  
\hline\end{array}
$$
\end{wraptable}
In this section we further explain the algebra of exchange-only control of triple-dot spin qubits.
This qubit is defined in a subsystem of angular-momentum states whose energy differences are immune to fluctuations in global magnetic fields \cite{DiVincenzo2000}.  
If we notate the total spin of the three electrons as $\vec{S}=\vec{S}_1+\vec{S}_2+\vec{S}_3$ and the total spin of the first two electrons as $\vec{S}_{12}=\vec{S}_1+\vec{S}_2$, then by standard rules for the addition of angular momentum, all 8 spin-states may be written as $\ket{S_{12},S;m}$, where $S_{12}$ takes on the values 0 or 1 and $S$ takes on the values $1/2$ and $3/2$.   
Our qubit states are defined as
\begin{equation}
\begin{aligned}
\ket{0}\otimes\ket{m}_{1/2}&=\ket{0,1/2;m},\\
\ket{1}\otimes\ket{m}_{1/2}&=\ket{1,1/2;m},\\
\ket{Q}\otimes\ket{m}_{3/2}&=\ket{1,3/2;m}.
\end{aligned}
\end{equation}
Although $m$ is a collective property of all three spins, we treat it as a separate subsystem which we do not initialize, measure, or control.  Critically, $m$ is the only degree of freedom coupled to global magnetic fields, and so ignoring $m$ is tantamount to ignoring all global magnetic fields, including global fields which may be fluctuating in any vector direction\footnote{Note that the two encoded states $\ket{0}$ and $\ket{1}$ are each associated with two values of $m=\pm 1/2$ and the leaked state $\ket{Q}$ is associated with four $\ket{m}$ states $m=-3/2,\ldots,3/2;$ here $Q$ stands for quadruplet.   The varying number of $\ket{m}$ states means the subsystem separation is imperfect; a true tensor product of subsystems could be recovered by treating the $\ket{m}$ subspace as four-dimensional and imposing a selection rule preventing population of $\ket{S_{12},1/2;\pm 3/2}$ states.}.
Our initial state may be written as $\ket{0}\!\!\bra{0}\otimes\rho_m$, where $\rho_m$ is a fully mixed (that is, completely unknown) state for the $m$ degree of freedom.  
Likewise, the singlet measurement operator is the projector
\begin{equation}
P_0 = \ket{0}\!\!\bra{0}\otimes I_m = \sum_{m=\pm 1/2} \ket{0,1/2;m}\!\!\bra{0,1/2;m}.
\end{equation}

The matrix elements for Heisenberg exchange between any two spins in this basis can be expressed via the Wigner-Eckart theorem \cite{messiah1962}, which first of all indicates that exchange leaves the quantum numbers $S$ and $m$ invariant, second that $\bra{S_{12},3/2;m}\vec{S}_j\cdot\vec{S}_k\ket{S_{12}',3/2;m}=1/4$ for any spin pair $j,k$ (and hence leaked states are not affected by exchange), and finally that the matrix elements for the qubit states in terms of Wigner 6$j$ coefficients (notated $\{\cdot\}$) and reduced matrix elements (notated $\bra{}\!|\cdot|\!\ket{}$) are given by
\begin{align}
\hspace{0.2in}
\bra{S_{12},1/2;m}\vec{S}_1\cdot\vec{S}_2\ket{S_{12}',1/2;m}
    &=\delta_{S_{12}S_{12}'}(-1)^{S_{12}+1}
    \left\{\begin{array}{ccc}1/2 & 1 & 1/2\\
                             1/2 & S_{12} & 1/2\end{array}\right\}\bra{1/2}\!|\hat{\vec{S}}|\!\ket{1/2}^2 \notag\\
     &=-\frac{1}{4}-\frac{1}{2}\sigma^z,
     \\
\hspace{0.2in}
\bra{S_{12},1/2;m}\vec{S}_2\cdot\vec{S}_3\ket{S_{12}',1/2;m}
    &=(-1)^{1+S_{12}+S_{12}'}\sqrt{(2S_{12}+1)(2S_{12}'+1)}
    		\left\{\begin{array}{ccc}S_{12} & 1 & S_{12}' \\
                                       1/2  & 1/2 & 1/2\end{array}\right\}^2
                                \bra{1/2}\!|\hat{\vec{S}}|\!\ket{1/2}^2 
    \notag\\
    &= -\frac{1}{4}+\frac{1}{4}\sigma^z+\frac{\sqrt{3}}{4}\sigma^x,
\end{align}
where $\sigma^x$ and $\sigma^z$ are $2\times 2$ Pauli matrices in the basis of the $\ket{0}$ and $\ket{1}$ states.  
Neglecting overall phase, we therefore define the effective exchange-only Hamiltonian
\begin{equation}
H_{\text{exchange}}(t) = J_z(t)\vec{S}_1\cdot\vec{S}_2
                        +J_n(t)\vec{S}_2\cdot\vec{S}_3
        \rightarrow-\frac{1}{2}J_z(t)\sigma^z -\frac{1}{2}J_n(t)[\sigma^z\cos\phi-\sigma^x\sin\phi],
\end{equation}
where $\phi=2\pi/3.$  
Hence in the $\ket{0}/\ket{1}$ Bloch sphere, exchange between dots 1 and 2 causes rotations about the $z$-axis, and exchange between dots 2 and 3 causes rotations about the $n$-axis, which is rotated from the $z$-axis in a right-handed sense about the $y$-axis by 120$^\circ$.

Qubit control is accomplished using generalized Euler angles by pulsing a single axis at a time.  
Table~\ref{table:Clifford_angles} shows all of the angles used to generate the 24 single-qubit Cliffords.
As an example, consider the Hadamard gate $H$, which is constructed via a 3-pulse sequence.
The middle pulse turns on $J_n(t)$ at a level to drive a rotation of angle $\int_0^\tau J_n(t) dt=\pi+\tan^{-1}\sqrt{8}$ in pulse time $\tau$.  
This angle is chosen as the one which rotates the $z$-axis to the $xy$ plane.  
Additional $z$-rotations convert this rotation to a Hadamard; our composite gate therefore begins and ends by pulsing $J_z$ for a total angle of $\int_0^\tau J_z(t) dt = \tan^{-1}\sqrt{2}$.  
The other Cliffords are constructed similarly; note that these constructions are not unique, but are guaranteed to exist and to require no more than four pulses for any single-qubit gate~\cite{lowenthal1971}.

Local magnetic field gradients, most typically caused by hyperfine interactions with $^{29}$Si nuclear spins, cause leakage from the $\ket{0}$ and $\ket{1}$ states into the $\ket{Q}$ states. 
They are therefore not describable within the Bloch sphere picture.  
For magnetic fields sufficiently high to ignore electron-nuclear flip-flop terms of the contact hyperfine interaction, an SU(3) picture is helpful to quantify leaking during and between pulses~\cite{ladd2012}.  
Note that by the addition rules for angular momentum, $S_{12}=1$ for all $S=3/2$ leaked states, and therefore all leaked states appear as triplet states on spins 1 and 2, and are measured as such in spin-blockade.

\section{Details of Rotation Calibration}
\label{sec:calibration_supplement}
The purpose of calibration is to generate a one-to-one mapping between exchange rotation and voltage throw for symmetric operation of the respective axis \cite{reed2016}.
To formulate this mapping, we apply a two-stage procedure consisting of `rough' and `fine' calibration stages.

\begin{figure}[h]
\includegraphics[width=0.8\linewidth]{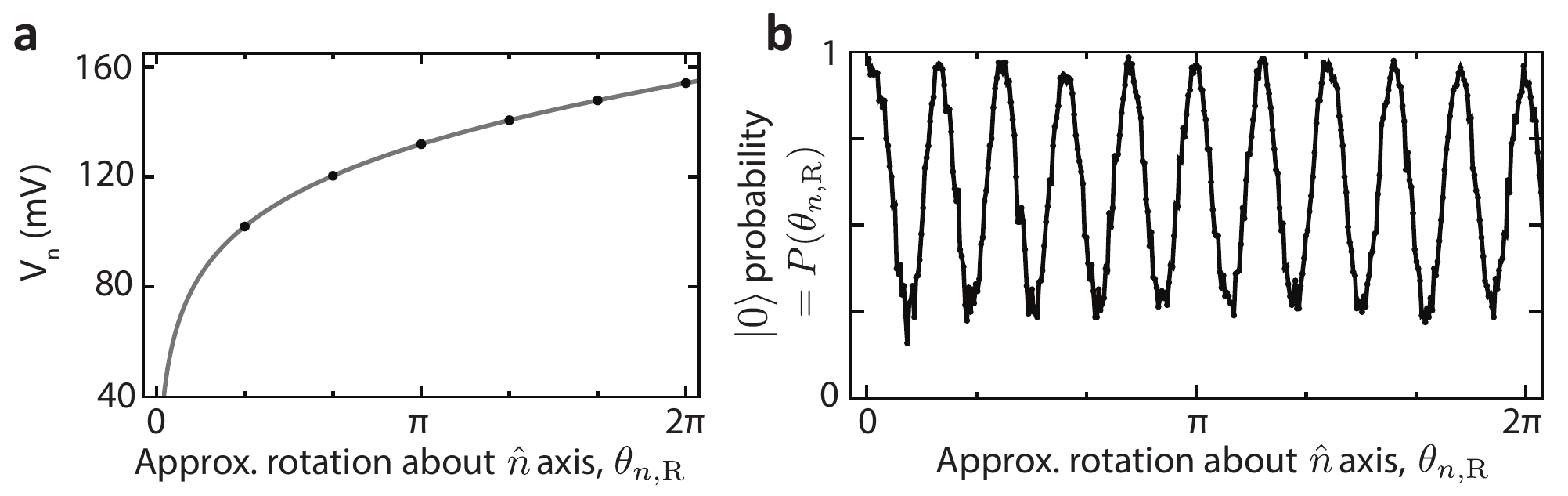}
\caption{ 
Details of axis calibration.  
{\bf{a}}, Rough mapping between rotation angle and applied voltage.  
{\bf{b}} Data used for `fine' calibration showing state probability as a function of rough rotation angle ($\theta_{n,\mathrm{R}}$).  
Note that this signal is very nearly monochromatic.}
\label{roughcal}
\end{figure}

In the `rough' calibration step, we estimate a functional form of $f_\theta = \theta_{\mathrm{\sigma}}(V_{\mathrm{\sigma}})$ that relates applied voltage $V_{\sigma}$ to rotation $\theta_{\sigma}$ about the $\sigma=n,z$ axes.
As described in Fig.~2b of the main text, we throw three consecutive exchange pulses with equal voltage amplitudes chosen to be along a vector $\vec{V}$ which has been calibrated to account for capacitive cross-talk of neighboring gates.
This `symmetric operation tangent' primarily adjusts the height of the potential barrier between electrons while keeping their chemical potential fixed.
A full rough calibration (shown in Fig.~2b of the main text) is acquired by performing numerous experiments whose amplitudes are chosen to be equally spaced along the symmetric operation tangent.
We find initial estimates of coordinates $\{V,\theta\}_{\mathrm{R}}$ by mapping the turning points of $|0\rangle$ probability to integer multiples of $\pi$.
The subscript `R' indicates that this is a rough estimate of the voltage-exchange pair.
We initially fit these seed coordinates to a function of the form
\begin{align}
\label{wkb}
f_\theta = V_{n}(\theta_{n,\mathrm{R}})=-2A\times \mathrm{Ln}\left[ \frac{\theta_{\mathrm{max}}-\theta_{n,\mathrm{R}}}{\theta_{\mathrm{max}}\sqrt{\theta_{n,\mathrm{R}}}}  \right]+B
\end{align}
where $A$ and $B$ are constants in units of voltage, and $\theta_{\mathrm{max}}$ is the asymptotic maximum of rotation angle. 
For the data shown in Fig.~2b of the main text, $A=24.24$, $B=59.29$, and $\theta_{\mathrm{max}}=32.96 $; the resulting curve is shown in Fig.~\ref{roughcal}a.

In the `fine' calibration step, we apply a series of 10 pulses at voltages that we now linearly sample in the `rough' calibration domain specified by Eqn.~\ref{wkb}.
That is, we resample the voltages along the symmetric operating tangent as $V_n = f_\theta(\theta_i)$ where $\theta_i$ is linearly sampled over the user-specified range $[\theta_{\rm min}, \theta_{\rm max}]$.
The resulting narrow-band signal appears nearly monochromatic (Fig.~\ref{roughcal}b) and the corrections needed to create an accurate calibration are encoded as deviations from a single-frequency oscillation.
Because of this, we can strongly reject fluctuations in the data that occur at both high and low frequencies as noise.  
We analyze the data by first extending it using Burg's maximum entropy method \cite{burg1975}, and then apply a FIR filter with a passband of 0.5 to 2 times the signal frequency.  

To estimate phase, we construct the analytic signal of the filtered, smoothed signal via the Hilbert transform and calculate the instantaneous phase as
\begin{align}
\theta_{n}(V_n) = \frac{1}{N_p}\mathrm{ArcTan} \left[ \mathcal{H} \left( \widetilde{P}(V_n) \right),   \widetilde{P}(V_n) \right],
\end{align}
where $\mathrm{ArcTan}()$ is the four-quadrant inverse tangent, $N_p=10$ is the number of pulses, $\mathcal{H}$ is the Hilbert transform and $\widetilde{P}(V_n)$ is the filtered and extended data. 
We remove the extension data introduced by Burg's method and unwrap the angles to produce a set of voltage-exchange pairs $\{V_n,\theta_n\}$ at every sampled voltage.
For proper unwrapping of the angles, we sample at sufficiently low exchange rates to ensure that the first point falls within the branch $\left[ 0, 2\pi/N_p \right]$.
Finally, we convert the mesh of voltage-exchange pairs into a continuous function via nonlinear interpolation with Eqn.~\ref{wkb},
\begin{align}
V(\theta) = f_\theta \left[  \left( 1-\alpha \right) f_\theta^{-1} \left( V_i \right) + \alpha f_\theta^{-1} \left( V_j \right) \right],
\end{align}
where $\alpha = (\theta-\theta_i)/(\theta_j - \theta_i)$ and $\{ V_i, \theta_i \}$, $\{ V_j,\theta_j \}$ bound the interval containing the specified $\theta$.

\section{Derivation of Blind RB}

Blind RB modifies the conventional RB experiment~\cite{knill2008} by estimating the probability that the system has leaked out of the qubit subsystem.
Whereas traditional RB only analyzes the population for a two-state system ($\ket{0}$ and $\ket{1}$), Blind RB estimates the population for a third state as well (labeled $\ket{Q}$ in our exchange-only qubit).
As an aside, there can be multiple leaked states, and this analysis considers their collective population.
The intuition behind Blind RB can be seen in projectors: let $P_0$, $P_1$, and $P_L$ be the projectors onto $\ket{0}$, $\ket{1}$, and $\ket{Q}$ (leaked states), respectively.  
Recall from the main text, and especially Fig.~1c, that $P_0$ and $P_1$ are two-dimensional because of the undetermined $m$ quantum number, while $P_L$ is four-dimensional (all four of the $S = 3/2$ states are leaked).
Since $P_L$ spans all of the states outside of the qubit basis, we have the normalization $P_0 + P_1 + P_L = \identity$.
The spin-to-charge measurement reports the population in $\ket{0}$ for any density matrix $\rho$ as $\mathrm{Tr}\left(P_0 \rho\right)$, assuming no measurement errors.
If we can also measure the population in $\ket{1}$, then we can determine the leaked population as $\mathrm{Tr}\left(P_L \rho\right) = 1 - \left[\mathrm{Tr}\left(P_0 \rho\right) + \mathrm{Tr}\left(P_1 \rho\right)\right]$.  
This intuition assumes no state-preparation-and-measurement (SPAM) errors; however, we show below how Blind RB is able to separate gate errors from SPAM errors in the same manner as conventional RB~\cite{knill2008}.

There is a straightforward way to measure $P_1$ if you have a measurement process for $P_0$ and control of the qubit: apply an $X$ gate just before measurement.
For some state $\rho$, this simply leads to a measurement probability
\begin{equation}
\mathrm{Tr}\left(P_0 X \rho X^{\dagger}\right) = \mathrm{Tr}\left(X^{\dagger} P_0 X \rho \right) = \mathrm{Tr}\left(P_1 \rho \right),
\end{equation}
where we have used the cyclic property of the trace.
However, to measure both $P_0$ and $P_1$, one must have multiple copies of $\rho$.  
This is possible in randomized benchmarking because $\rho$ is an ensemble over all Clifford circuits of some length $N$, which is approximated with some randomly sampled subset of such Clifford circuits.
Hence we can apply the technique above by preparing many copies of $\rho(N)$, the state after a random $N$-Clifford circuit, and measuring some with $P_0$ and the others with an $X$ gate followed by $P_0$, to approximate $P_1$.

Blind RB makes one additional modification to measuring $P_1$ effectively.
Since gates are not perfect, one might be concerned that any experiment that attempts to measure $P_1$ has slightly more error than measurement of $P_0$ due to any error in the added $X$ gate.
Blind RB addresses this by modifying the final ``recovery'' Clifford gate: half of the Clifford sequences compose to identity (target final state is $\ket{0}$), and the other half compose to $X$ (target final state is $\ket{1}$).
Because the $X$ gate is compiled into the recovery operation, one can measure $P_1$ in this circumstance without the additional gate error.

Having established the intuition for Blind RB, we now show more formally how the experiment is analyzed.  Recall from the main text that the probabilities for measuring the target final state are binned in two groups $y_0(N)$ and $y_1(N)$, where subscript is the target state, as a function of sequence length $N$.  If $\rho(N)$ is the state after $N$ Cliffords, then the state after $N$ Cliffords with the flipped recovery operation is approximately $X \rho(N) X$, and the measured probabilities are
\begin{align}
& y_0 = \mathrm{Tr}\left[P_0 \rho(N)\right], \\
& y_1 = \mathrm{Tr}\left[P_0 X \rho(N) X\right] \approx \mathrm{Tr}\left[P_1 \rho(N)\right].
\end{align}
We can begin to analyze the results by first considering leakage.
The sum combination can be expressed using the normalization as
\begin{equation}
W(N) = y_0(N) + y_1(N) \approx \mathrm{Tr}\left[\left(P_0 + P_1\right) \rho(N)\right] = 1 - \mathrm{Tr}\left[P_L \rho(N)\right],
\end{equation}
where approximate equality accounts for the presence of SPAM errors.  We fit the measured $W(N)$ to the ansatz
\begin{equation}
W(N) = 2A + 2C (1-q)^N,
\end{equation}
where parameters $A$ and $C$ account for the unknown number of accessible leaked states and SPAM error.  We conjecture that $W(N)$ decays as a single exponential in $N$.  The intuition for this is that Clifford twirling in the qubit subsystem will interrupt any coherent driving between qubit and leaked states, emulating a $T_1$-like process.  This conjecture is supported by the data in Fig.~3d of the main text showing that an exponential fit follows the data closely.  More detailed theory for this matter is the subject of future work.

We can also extract errors within the qubit subystem by fitting to the difference combination.
Let $R(N) = y_0(N) - y_1(N)$: 
\begin{equation}
R(N) \approx \mathrm{Tr}\left[\left(P_0 - P_1\right)\rho(N)\right],
\end{equation}
where again approximate equality accounts for the presence of SPAM errors.  We fit this to
\begin{equation}
R(N) = 2B (1-p)^N.
\end{equation}
Other works have discussed the advantage of the $(P_0 - P_1)$ ansatz for fitting~\cite{muhonen2015,helsen2017}, but we see that it is also advantageous in the presence of leakage since $P_0 - P_1 = \identity - (2 P_1 + P_L)$.  
By randomizing what the final state of the experiment should be, a leakage event has a 50\% chance of being correct, producing the same measurement statistics in $y_0 - y_1$ as a depolarizing error within the qubit subspace.  
As a remark, this is where ``blind'' comes from in the name of the protocol -- during any experiment, the system is blind to whether the correct final answer should be $\ket{0}$ or $\ket{1}$ without knowing all of the Clifford gates in the sequence.
Combined with standard RB assumptions for Clifford twirling of gate errors, $R(N)$ is expected to be a single exponential decay.

Echoing the main text, the total error per gate and leakage per gate are defined in the following way, using the analysis methods of randomized benchmarking to separate out SPAM errors~\cite{knill2008}.  Total error $\varepsilon$ is the population leaving $\ket{0}$ in first gate, normalized by the ``visibility'' of the experiment $(2B)$:
\begin{equation}
\varepsilon = \mathrm{Tr}\left\{P_0[\rho(0) - \rho(1)]\right\}/(2B) = p/2 + Cq/(2B).
\end{equation}
As was noted in the main text, this definition will coincide with the gate error from conventional RB when there is no leakage.
Whereas conventional RB assumes the dimensionality of the qubit to be 2 by definition, we use the visibility $2B$ to normalize error, which is essentially the contrast between two experiments: (a) prepare $\ket{0}$, measure $P_0$ and (b) prepare $\ket{0}$, apply $X$, measure $P_0$.
Ideally, $2B = 1$, but it can be less due to SPAM error.
In the next paragraph, we examine the effect of SPAM errors more deeply, which helps to explain why we say that we ``normalize by SPAM''; what this analysis does is infer what the rate of population transfer would be (for gate error and for leakage) if there were no SPAM errors.
One can apply the same analysis to estimate fidelity as
\begin{equation}
F_{\mathrm{SPAM}} = 0.5 + B,
\end{equation}
which we derive below.
Similar to total error, leakage $\Gamma$ is the population leaving the qubit subsystem in the first gate, also normalized by $2B$:
\begin{equation}
\Gamma = \mathrm{Tr}\left\{[P_L[\rho(0) - \rho(1)]\right\}/(2B) = Cq/B.
\end{equation}
We use the term ``qubit error'' (denoted here as $\epsilon_q$) to correspond to the probability of error per gate that is not leakage outside of the encoded qubit, which is simply $\epsilon_q = \varepsilon - \Gamma$.

The fits also enable estimating the population in each of the states $\ket{0}$, $\ket{1}$, and $\ket{Q}$ during the course of the Blind RB experiment.  
These state probabilities are shown in Fig.~3c and Fig.~4a of the main text.  We must also account for SPAM, so we introduce two more ``states,'' $M_0$ and $M_1$.  
These are the probabilities that the final measurement will report ``0'' or ``1'' regardless of the expected state of the triplet-dot system, indicating error during preparation, measurement, or both.  
As they are SPAM errors, $M_0$ and $M_1$ are not functions of the number of Clifford gates $N$, by assumption.  
In general, $M_0 \neq M_1$ if there is a bias to one outcome in the measurement process (this can occur due to T1 decay during measurement~\cite{eng2015,jones2018}, among other reasons).  
Let the following be probabilities for actually measuring the correct state of the system, meaning no SPAM error:
\begin{align}
& \nu_0 \equiv \textrm{Pr(no SPAM error)}\mathrm{Tr}(P_0 \rho); \\
& \nu_1 \equiv \textrm{Pr(no SPAM error)}\mathrm{Tr}(P_1 \rho); \\
& \nu_L \equiv \textrm{Pr(no SPAM error)}\mathrm{Tr}(P_L \rho).
\end{align}
Since $M_0+M_1 = \textrm{Pr(SPAM error)}$, we have the normalization
\begin{equation}
\nu_0 + \nu_1 + \nu_L + M_0 + M_1 = 1.
\end{equation}
Based on these definitions, we can relate these terms to the fits to $y_0$ and $y_1$ as follows:
\begin{align}
y_0(N) = A + B(1-p)^N + C(1-q)^N = \nu_0(N) + M_0; \\
y_1(N) = A - B(1-p)^N + C(1-q)^N = \nu_1(N) + M_0.
\end{align}
By assumption, the qubit is ideally initialized as $\ket{0}$, so $\nu_1(0) = \nu_L(0) = 0$.  
Since we assume $M_0$ and $M_1$ are independent of $N$, we can solve for them at $N = 0$ and produce the following probabilities for all $N$:
\begin{align}
& M_0 = A - B + C; \\
& \nu_0(N) = y_0(N) - M_0 = B\left[1 + (1-p)^N\right] + C\left[(1-q)^N - 1\right]; \\
& \nu_1(N) = y_1(N) - M_0 = B\left[1 - (1-p)^N\right] + C\left[(1-q)^N - 1\right]; \\
& M_1 = 1 - y_0(0) = 1 - (A + B + C); \\
& \nu_L(N) = 1 - \left[M_0 + M_1 + \nu_0(N) + \nu_1(N)\right] = 2C\left[1 - (1-q)^N\right].
\end{align}
From this, we can derive SPAM fidelity as
\begin{equation}
F_{\mathrm{SPAM}} = 1 - 0.5\left(M_0 + M_1\right) = 0.5 + B,
\end{equation}
where the factor $1/2$ comes from the fact that, averaging over all states, the errors where the measurement returns ``0'' or ``1'' regardless of input will be correct half of the time (this follows the convention where single-qubit RB error is half of the equivalent depolarizing-channel parameter~\cite{knill2008}).
In the main text, we plot an additional curve in each of Fig.~3c and Fig.~4a, which is $y_0(N) + \nu_1(N)$.
As indicated by the vertical arrows and described in the caption, the area between this curve and $y_0(N)$ is the estimated probability for the qubit be in $\ket{1}$, as a function of sequence length $N$.  
Likewise, the area between this curve and above to $(1 - M_1)$ is $\nu_L(N)$, the probability of leakage.
However, since $M_0$ and $M_1$ are each about 0.4\%, we excluded them from the plots since they would be imperceptible.

Finally, we make a few remarks on the analysis. 
One can say that the parameter $q$ or the combination $y_0 + y_1$ measures leakage because of the proportionality $\Gamma \propto q$.  
However, it is not correct to say that parameter $p$, nor the combination $y_0 - y_1$, measures total error or qubit error.
This can be seen in the definition of total error, which can restated as $\varepsilon = p/2 + \Gamma /2$.
In fact, the rate of exponential decay in $y_0(N) - y_1(N)$ lies exactly halfway between twice the total error and twice the qubit error:
\begin{equation}
p = 2 \varepsilon - \Gamma = 2 \epsilon_q + \Gamma.
\end{equation}
The reason for this connects back to the discussion above about why Blind RB is called ``blind.''
The decay seen in $R(N) = y_0(N) - y_1(N)$ accounts for every qubit error, but only half of the leak errors.
Mathematically, this is captured in the expression above that $P_0 - P_1 = \identity - (2 P_1 + P_L)$.
Alternatively, one can say that every qubit error will register as incorrect, while a leakage error will be correct half of the time.
In the exchange-only qubit, a leaked state will always register as the Pauli blockaded state (i.e. ``not $\ket{0}$'') when measured; when the correct final state is randomized, this faulty measurement will be right 50\% of the time.
In some contexts, it may indeed be the case that a leaked qubit can be treated as if it were depolarized; for example, the leakage-reduction unit (LRU) derived in Ref.~\cite{fong2011} will turn a leaked exchange-only qubit into a non-leaked one, but its state relative to what it ideally would be is effectively depolarized.
Assuming this LRU will be applied at some point, one could report the error per gate as simply $p/2$.
However, we have chosen to take the conservative approach and count all of leakage probability in our reported total error.

We also note that the way Blind RB looks at Clifford sequences that should compose to identity or $X$ (i.e. end in $\ket{0}$ or $\ket{1}$, respectively) is very similar to the method in Ref.~\cite{wallman2016}, which proposes to simply omit the final inversion gate of an RB experiment.
By leaving out the inversion gate, 1/6 of the sequences will end ideally as $\ket{0}$, 1/6 will end as $\ket{1}$, and the remaining 2/3 will end with 1/6 probability each in the $\ket{\pm}$, $\ket{\pm i}$ eigenstates of the $X$ and $Y$ operators.
Measuring $P_0$ on this mixed state will give the same expectation as $(y_0 + y_1)/2$, but the variance will be higher for the same number of experimental samples due to measurement-projection noise from the $\ket{\pm}$ and $\ket{\pm i}$ states.
Simply put, Blind RB focuses on the Clifford sequences that are more informative, giving more accurate estimates of leakage error for the same number of experimental runs.

\section{Error Scaling with Overrotation}

\newcommand{\abs}[1]{\left\vert #1 \right\vert}

To obtain the data of Fig. 4c of the main text, we first calibrate the $\hat{n}$ and $\hat{z}$ axes following the calibration procedure described in Sec.~\ref{sec:calibration_supplement}. 
The voltage-to-rotation angle mapping, $\theta(V)$, is then used to select rotations which differ from the nominal value by a percentage $\Delta$, such that $\theta_i \rightarrow (1 \pm \Delta) \theta_i$.
The sign of the overrotation is randomly chosen prior to running the blind randomized benchmarking routine.
In Fig. 4c, multiple data points for a given value of $\Delta$ represent distinct instantiations of the randomly-selected sign of the overrotation.

From the definition of gate fidelity for unitary operators \cite{nielsen2002},
\begin{equation}
\label{gateFidelity}
F_g\left(U,\tilde{U}\right) = \frac{\abs{\textrm{Tr}\left(\tilde{U}^{\dagger} U\right)}^2/d+1}{d+1},
\end{equation}
we can calculate the effect of an implemented unitary, $\tilde{U}$, relative to the ideal, $U$, in a system of dimension $d$. 
For a given percent angular overrotation, $\Delta$, we have 
\begin{flalign}
\label{unitaries}
& U = \mathrm{exp}\left(-i\theta \sigma_{n}/2\right),\\
& \tilde{U} = \mathrm{exp}\left[-i(\theta+\Delta \theta)\sigma_n/2\right].
\end{flalign}
Inserting these into the trace of Eqn. \ref{gateFidelity} gives
\begin{equation}
\label{gateFidelity2}
\abs{\textrm{Tr}\left(\tilde{U}^{\dagger} U\right)}^2 = \abs{\textrm{Tr}\left[\mathrm{exp}\left(i\Delta \theta \sigma_n/2\right)\right]}^2 = \abs{\textrm{Tr}\left[\textrm{cos}\left(\Delta\theta/2\right)\mathbb{I}-i~\textrm{sin}\left(\Delta \theta/2\right)\sigma_n\right]}^2 = 4~\mathrm{cos}^2(\Delta \theta/2).
\end{equation}
Taking $d=2$ and performing a linear expansion of the infidelity, $1-F_g(U,\tilde{U})$, we obtain the expected scaling of a single gate error from overrotation
\begin{flalign}
\label{gateError}
\varepsilon_g = 1-F_g\left(U,\tilde{U}\right) = \frac{2}{3}\mathrm{sin}^2(\Delta \theta/2) \approx \frac{(\Delta \theta)^2}{6} + \mathcal{O}(\Delta \theta)^4.
\end{flalign}
For an average rotation angle of $\theta = \pi$ and 2.7 pulses per Clifford, this yields an expected scaling of $2.7\frac{\pi^2}{6}\Delta^2 \approx 4.4 \Delta^2$.  In the main text, Fig.~4c, we show that the total error from Blind RB scales as $0.0042 + 4.3\Delta^2$, where the constant offset reflects the intrinsic qubit error in the absence of intentional overrotation.

\section{Simulations of Randomized Benchmarking Experiments}
\label{sec:simulation}
The simulations in Fig. 4d of the main text mimic the experimental procedure quite closely, with accurate models for magnetic and charge noise.
The primary differences are that the simulation uses perfect initialization and measurement, and exchange control is implemented as square pulses with noise on exchange rates $J_z(t)$ and $J_n(t)$ but without the pulse distortion from bandwidth limitations that we believe to be present.
The contact-hyperfine interactions between electron spins and nuclei ($^{29}$Si and $^{73}$Ge) are approximated as a fluctuating Overhauser field for each electron spin.
The simulation uses the full 8-dimensional basis of 3 spins, allowing arbitrary pairwise exchange and local magnetic field vectors on each spin.  
The simulated experiment is as follows: 
\begin{enumerate}
\item Preparation -- a singlet on two spins is prepared with a third spin with random orientation.
\item Evolve -- a blind RB pulse sequence proceeds using the same Clifford constructions as the experiment, with each exchange rotation implemented as a square pulse with duration of 10~ns followed an by idle ($J = 0$) duration $t_{\mathrm{idle}}$ as indicated for the experiment or 20~ns when not specified.  The effects of the magnetic environment are simulated during all pulse and idle intervals.
\item Measurement -- the two-spin singlet probability is directly extracted from the simulated wavefunction.  
\end{enumerate}
The simulated Hamiltonian has the following form:
\begin{equation}
H(t) = J_z(t)\vec{S}_1 \cdot \vec{S}_2 + J_n(t)\vec{S}_2 \cdot \vec{S}_3 + \sum_{k = 1}^3 \mathbf{b}_k(t) \cdot \vec{S}_k,
\label{eqn::sim_Hamiltonian}
\end{equation}
where each $\mathbf{b}_k(t)$ is the simulated Overhauser field for dot $k$.
Noise on each $J_\sigma(t)$ and $\vec{b}_k(t)$ is simulated as described below.
The simulator approximates evolution with a first-order Magnus expansion:
\begin{equation}
\mathcal{T}\mathrm{exp}\left(-\frac{i}{\hbar}\int_{t_a}^{t_b} H(t)dt\right) \approx \mathrm{exp}\left(-\frac{i}{\hbar}\int_{t_a}^{t_b} H(t)dt\right).
\end{equation}
Each simulation interval $[t_a,t_b)$ is a single pulse or idle interval between pulses.
After a single simulated experiment, the simulator is advanced forward in time equal to the experimental timescale for repeating the same pulse sequence and averaging the single-shot measurements to estimate the population of $\ket{0}$.  
In order to capture any long-timescale correlations due to low-frequency noise, the schedule of performing RB with multiple random Clifford sequences and multiple averages closely mimics the schedule in the actual experiment.

\begin{figure}[h]
\includegraphics[width=1\linewidth]{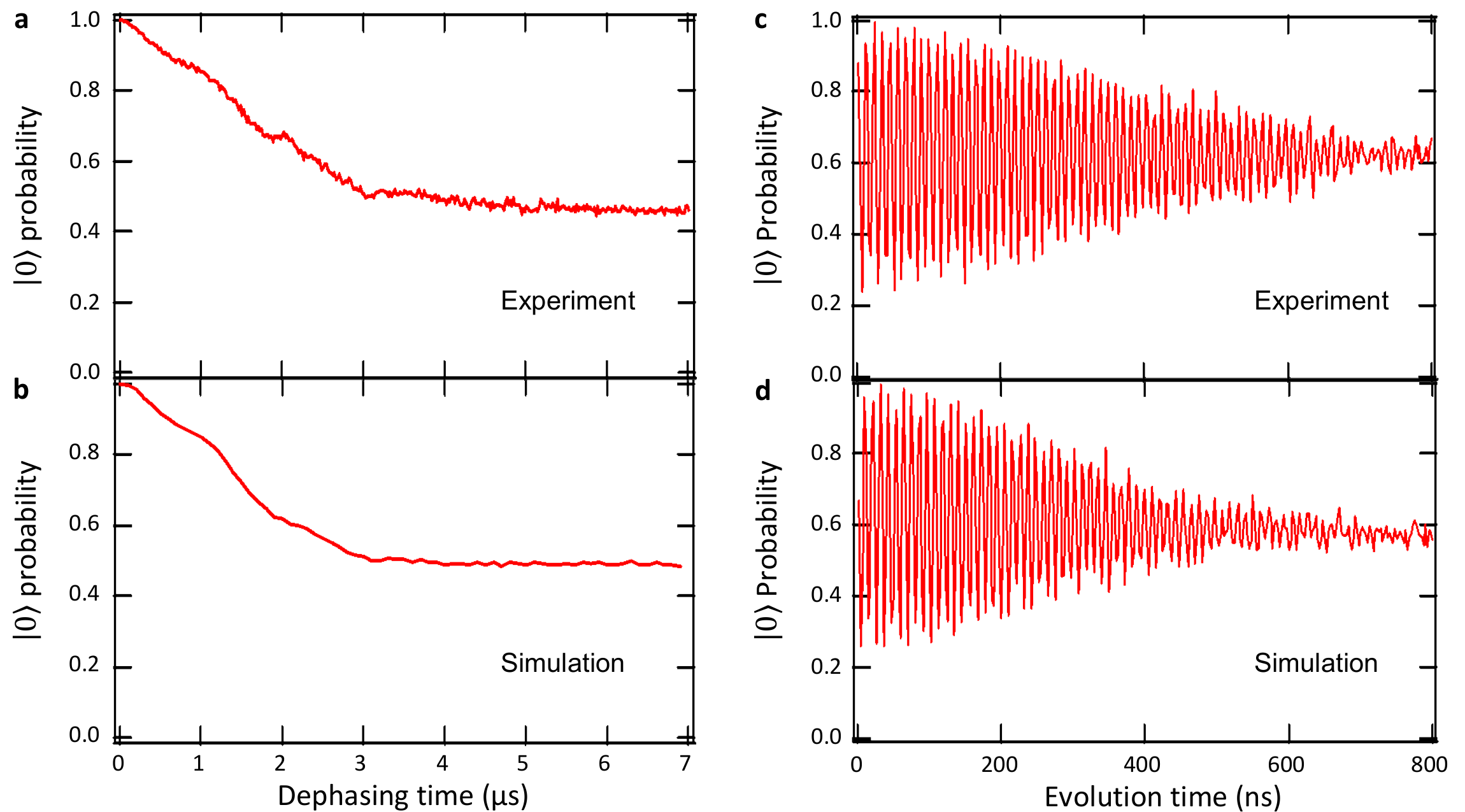}
\caption{
Measurement of singlet dephasing due to charge noise and fluctuating hyperfine-induced magnetic field gradients at low field.  A singlet is prepared on dots 1 and 2, allowed to dephase in the presence of {\bf{a}}, negligible exchange, or {\bf{c}}, 93 MHz exchange between dots 2 and 3, and then measured. {\bf{b}}, Simulation of the experiment in {\bf{a}}.  {\bf{d}}, Simulation of the experiment in {\bf{c}}.}
\label{fig:FID}
\end{figure}

Noise is introduced in two forms, the Overhauser field that approximates the impact of contact-hyperfine interactions with nuclei and charge noise affecting the exchange interaction.
Noise with $1/f$-like power spectrum is simulated through a variant of the Voss-McCartney algorithm, which is essentially an ensemble of gaussian fluctuators with appropriately chosen switching rates, extended to evolution over arbitrary-length intervals.
First, the magnetic noise is an independent $1/f$ ensemble for each of the $b^x_k(t)$, $b^y_k(t)$, $b^z_k(t)$ magnetic fields for each electron.
The $1/f$ noise spectral density corresponds to observed magnetic noise in this and previous Si/SiGe quantum dot devices~\cite{eng2015}.  
The spectrum is bounded by providing a low-frequency cut-off to white noise below 1~mHz   and a high-frequency roll-off to $1/f^2$ for noise above 100~kHz.  
These timescales are approximately based on experimentally observed averaging times~\cite{eng2015} and bounds to the hyperfine and nuclear dipole-dipole Hamiltonian.   
Simulation results depend very weakly on the lower- and upper-frequency roll-offs.  
The total integrated noise is chosen to match the observed degree of dephasing observed in the decay of double-dot singlet coherence in this device (shown in Fig.~\ref{fig:FID}a), which is similar to that discussed in Ref.~\onlinecite{eng2015}.   
Note that the decay shape and asymptotic singlet probability in Fig.~\ref{fig:FID}a match expectations for a singlet dephasing under slowly varying, independently fluctuating magnetic fields in double dots at low applied magnetic field~\cite{schulten1978}; in this case the magnetic field is that of the Earth ($\sim$50~$\mu$T). The corresponding simulation is shown in Fig.~\ref{fig:FID}b.

The second form of simulated noise is charge noise.  
Each exchange term in Eqn.~(\ref{eqn::sim_Hamiltonian}) is simulated as $J(t) = J_0\left[1 + \epsilon(t)\right]$, where $\epsilon(t)$ is a $1/f$ noise process having a high-frequency roll-off at 10~GHz, much faster than any pulsing time, and a low-frequency cut-off slower than any experimental timescale (typically mHz).  
For each pulse of angle $\theta$ and duration $t_{\mathrm{pulse}}$, $J_0 = \theta/t_{\mathrm{pulse}}$, i.e. the exchange needed for a square pulse to implement the intended rotation.
The simulator sets $J_0 = 0$ during idle periods, and only one exchange term is nonzero during pulsing; however, the simulated noise processes are advanced in time to properly include time correlations.
Notably, simulating square pulses and $J = 0$ during idle periods does not account for pulse distortion, which we consider to be an important noise source at short idle time, as discussed in the main text.
The amplitude of this $1/f$ spectrum is similar to that discussed in Ref.~\cite{reed2016}, consistent with Rabi-oscillation measurements in this device as shown in Fig.~\ref{fig:FID}c and providing the simulated decay of coherence shown in Fig.~\ref{fig:FID}d.

\bibliography{BlindRB_supplement.bbl}